\documentclass[prd,aps,showpacs,twocolumn]{revtex4}
\usepackage{graphicx}

\begin{document}

\title{
Decay width of light quark hybrid meson from the lattice.}

\author{C. \surname{McNeile}}
\author{C. \surname{Michael}}

\affiliation{Theoretical Physics Division, Dept. Math. Sci., University
of Liverpool,  Liverpool L69 7ZL, UK.}

\collaboration{UKQCD Collaboration}

 \begin{abstract} 
 Lattice QCD with $N_f=2$ flavours of sea quark is used to explore the
spectrum and decay  of a $J^{PC}=1^{-+}$ spin-exotic hybrid meson.
 We test lattice determination of S-wave decay amplitudes at threshold
using $b_1 \to \pi \omega$ where  agreement with data is found.    We
find  a hybrid meson state  at 2.2(2) GeV with a partial width to $\pi
b_1$ of  400(120) MeV and to  $\pi f_1$ of 90(60) MeV.
\end{abstract}


\maketitle

\section{Introduction}

QCD has the potential to produce spin-exotic hybrid mesons. Lattice QCD 
is one of the most reliable ways to evaluate their properties and  mass
values have been reported. For a successful experimental study of such
states,  it is necessary that the total decay width is not too wide. The
strength of this  decay in which a gluonic excitation produces a
quark-antiquark pair is not  easy to estimate phenomenologically. For
the case of heavy-quark hybrid mesons,  lattice QCD has given guidance
on the predominant decay channel and the decay width. This
estimate~\cite{McNeile:2002az}  is of a width sufficiently narrow that
experimental study is feasible. Here we  address the issue of the decay
mechanism and associated widths for the  spin-exotic meson made of light
valence quarks.

Lattice QCD offers a first principles route to determine the spectrum 
of spin-exotic hybrid mesons. The first reported results, 
refs.~\cite{Lacock:1996vy,Lacock:1996ny,Bernard:1997ib,Hedditch:2005zf},
 used quenched lattices. There have been subsequent studies using 
lattices with dynamical quarks,
refs.~\cite{Lacock:1998be,Bernard:2003jd},  but then the issue of the
decay of the  hybrid meson has to be addressed directly. Indeed the MILC
group~\cite{Bernard:2003jd}  emphasise that they cannot easily
distinguish a  hybrid meson from  a two-body state such as $\pi b_1$ 
with the same quantum numbers. 

This is an important field to explore, since experimental results are
somewhat  inconclusive and there have been several candidate states
proposed,  see 
refs.~\cite{Alde:1988bv,Thompson:1997bs,Abele:1998gn,Adams:1998ff,Chung:1999we,Ivanov:2001rv,Dzierba:2005jg,Kuhn:2004en,Lu:2004yn}.

The study of hadronic decays from the lattice is not straightforward - 
see ref.~\cite{Michael:2005kw}. It is possible, however, to evaluate the
 appropriate hadronic matrix element from a lattice if the transition is
 approximately on-shell. Since we will explore S-wave decays at
threshold,  we test our lattice methods on a case which is known
experimentally, namely $b_1 \to \pi \omega$,  obtaining  agreement. 
For the case of a spin-exotic hybrid meson (here we focus mainly  on the
isovector $J^{PC}=1^{-+}$ meson labelled $\hat{\rho}$ where the `hat'
notation implies opposite C),  the S-wave decays to $\pi b_1$ and $\pi
f_1$ are explored here. We follow methods  generically similar to those
used  by us to study $\rho$ decay~\cite{McNeile:2002fh}. Indeed these
methods were  first employed~\cite{McNeile:2002az} in a study of hybrid
meson decay, where the valence quarks  are taken as very heavy (i.e.
static). Here we use light valence quarks - lighter than the strange
quark.

\section{Lattice methods}

In order to study spin-exotic hybrid mesons, it is inevitable that
non-local  operators have to be used to create (and destroy) the hybrid
meson - since  hybrid mesons with spin-exotic quantum numbers explicitly
cannot be made from quark and antiquark alone. The  gluonic component
can be incorporated either by using a closed colour  flux loop (eg.
clover-like) or by separating the quark and antiquark  sites  by a
combination of colour flux paths. Here we use that latter construction -
 as it was found to be effective in an earlier
study~\cite{Lacock:1996vy,Lacock:1996ny}.

In order to construct the hybrid meson sources, one needs either 
propagators from different spatial points~\cite{Lacock:1996vy}, or some 
more elaborate construction.   The hybrid meson is relatively heavy, so
the signal to noise ratio will be poor. This suggests that a spatial-volume
source  would be attractive. Such sources are achievable using
stochastic methods -  and were indeed employed in our study of $\rho$
decay~\cite{McNeile:2002fh}.  Compared to the case of the $\rho$ meson,
the hybrid meson is even more challenging.  We explored a variety of
different prescriptions, attempting to optimise the  signal/noise for
the two-point hybrid correlator at moderate $t$-separations.  

 \subsection{Stochastic method}

 For a given effort (namely number of inversions),  the stochastic noise
is reduced the more one dilutes (or thins - see
refs.~\cite{Foster:1998vw,Foley:2005ac}) the set of stochastic  sources
used - until with  sources at one colour-spin at one point, one has the
conventional  exact inversion.  Conversely, the more correlators one
evaluates, the more the statistical noise  inherent in the gauge
configurations is reduced. So clearly we have to compromise - and the 
balance point for a hybrid meson may be different from that used, for
example, for $\rho$ decay. 

 As an example, we measured the connected  meson  two-point
 correlator at $t=8$  from different methods using  200 gauge
configurations (lattice U355 of Table~\ref{tab1}).  The standard deviations over
gauges  were found to be for the $b_1$ and $a_0$ meson 0.0028 and
0.0125, respectively,  from the conventional method using  12
colour-spin  sources at one space-time site; 0.0015 and 0.0070 from
using the same conventional method with four space-time sources per
gauge configuration; while with  a stochastic spatial-volume source we
obtained 0.0013 and 0.0034. The stochastic method is superior,
especially for the $a_0$ meson,  although it involves more inversions as
the method  used a sequence of  sources, each  at  one time value and
one colour-spin value but all space. With 8 time values  selected per
gauge configuration, this involves 12$\times$8 inversions. The essential
step in obtaining a small stochastic error is  to minimise the number of
random numbers needed to evaluate the relevant correlators. This we
explain in more detail.

 With a random source, $\xi_i$, which is non-zero on  some subset of
colour-spin-space-time and satisfies  
$ \langle \xi^{*}_i \xi_j \rangle =  \delta_{ij}$;
  $  \langle \xi_i \xi_j \rangle  = 0 $ when averaged over different 
instances of the random numbers, then one solves the propagation from
this  source iteratively, as usual, $M_{ik} \phi_k= \xi_i$. Here we use 
Z2 noise in both real and imaginary parts. The basic
idea is then that  $\xi_k^* \phi_j$ is an unbiased  stochastic estimator
of the propagator $M^{-1}_{jk}$, albeit a rather noisy one, and can be
used  to construct mesonic correlators. A considerable decrease in the
noise comes from using the \lq one-end-trick \rq \ which
is~\cite{Foster:1998vw,McNeile:2002fh} to combine two  appropriately
related $\phi$'s to obtain the mesonic two-point correlator.  So if
$M_{ik} \phi^{\Gamma}_k = \Gamma_{ij} \xi_j$, then $\phi^{*\Gamma} 
\phi$ will, averaged over stochastic samples, automatically select a
meson  created by $\bar{q} \Gamma q$. An appropriate sum also needs to
be made over the sink (indices available  on $\phi^{\Gamma}$ and $\phi$)
 to destroy such a meson. If one uses random sources $\xi$ for each spin
component separately, one  has the appropriate $\phi^{\Gamma}$ available
for any independent $\gamma$  matrix or product of them. This is then an
efficient method to evaluate  all mesons created by $\bar{q} \Gamma q$,
and we presented some results for the $b_1$ and   $a_0$ two-point
functions above.

In order to extend this approach to create a hybrid meson, one needs to
create a source $H_{ij} \xi_j$ for each original source $\xi_i$ where 
$H_{ij}$ is a sum of a set of operators comprising spatial paths and 
$\gamma$-matrices which create the hybrid meson when combined as above.
In this work we only create the spin-exotic $J^{PC}=1^{-+}$ meson in 
one spin component.  We also   create a fuzzed source $F_{ij} \xi_j$ to
give us an independent mesonic creation operator. From $\phi^{*H} \phi$
and $\phi^{*H} \phi^{F}$, we are able to create  a hybrid meson in two
different ways (the latter has some unwanted mixing of  opposite C,
which can be projected out by building the sink operator with the
required C-value). We also need  an extended propagator to cope with the
pion emission in the decays we shall study.  Thus overall we need
$4\times12\times8$  inversions per gauge configuration. This relatively
heavy overhead is justified because with dynamical fermions, there are a
limited set  of gauge configurations.

 \subsection{Lattice configurations and two point correlators}

 We wish to use gauge configurations with dynamical quarks so that 
decay is physically allowed, as in the real world. It is sufficient for
this  study to have $N_f=2$ flavours of sea-quark, since this allows the
light quark  sector to be explored. We use clover-Wilson fermions and
the lattice data sets used are  described in Table~\ref{tab1}. They have different 
quark masses and spatial volumes to explore systematic effects.

 Because of the improved signal to noise offered by our stochastic
method, we   have improved determinations of the meson masses from
two-point  correlators. As well as local operators, we use
fuzzing~\cite{Lacock:1994qx}  with paths of length $f_1$ ($2a$ for U355
for compatibility with previous  work, $3a$ for C410) composed of fuzzed
links (5 iterations  of fuzzing with 2.5 \ straight + sum U-bends,
projected to SU(3)).   We have a $2 \times3$ matrix of correlators:
local or fuzzed at  the source and an additional larger scale fuzzing at
the sink with fuzzed links with 10 iterations of fuzzing combined to
length  $f_2$ ($4a$ for U355, $5a$  for C410). For the axial mesons
$b_1$ and $a_1$, we use  operators $\bar{q}\gamma_i \gamma_5 \gamma_4 q$
and  $\bar{q}\gamma_i \gamma_5  q$,  respectively, (and their fuzzed
extensions) to determine the correlators which are fitted to give  the
ground state masses listed in  Table~\ref{tab2}.

 
 We are also able to extract the spin-exotic hybrid mass from our 
two-point correlators. We use  a hybrid operator made  (as in
ref~\cite{Lacock:1996vy}) of U-shaped paths $P_i$ from fuzzed links and
combined with  $\gamma$-matrices to be in the $T_1^{-+}$ representation:
 $\bar{q}\epsilon_{ijk} P_i \gamma_j q$.
 We have at the source such an operator with sides of length $f_1$ and
also  the fuzzed version of this (with straight links of length $f_1$
combined with the  U-shaped ones). At the sink we have both of these
operators as well as an additional one  made of U-shaped paths with
length $f_2$ composed of more heavily fuzzed links. At the sink we take
care to give the fuzzed-hybrid operator a well defined  charge
conjugation. 
 We again have a $2 \times 3$ matrix of correlators which we fit to 
determine the energy eigenstates. Using the $t$-range 2-8, we obtain 
the mass values given (for the ground state) in Table~\ref{tab2}  from a
factorising fit to the  correlator matrix with ground state and 1
excited state. 
 The fit is illustrated in fig.~1 for U355.
 We can also use a  $2 \times 2$ submatrix of correlators in a
variational treatment, obtaining $m(\hat{\rho})a=1.38(12)$ (U355 from
$t$-values 5/4 with basis 3/2)  and 1.68(13) (C410 from $t$-values 4/3
with basis 3/2). These latter  values are formally  upper limits. They 
are consistent with the fitted values given in Table~\ref{tab2}.  

 To compare these mass determinations, we use $r_0$, obtaining
$m(\hat{\rho})r_0= 7.0(3)$ and $5.4(2)$ for U355, C410 respectively, 
where the errors are statistical only. There are also systematic errors
from the fits,  which can be  constrained partly by the variational
results quoted above, and which are at least as big as the statistical
errors.
 These fit values are statistically barely consistent:  it may  be that
finite volume effects are significant (C410 has a spatial extent 5/3
bigger than U355) or that finite lattice spacing effects are important 
(U355 has a smaller lattice spacing by 3/5 and uses NP clover rather
than tadpole-improved clover). 
 Since we are unable to distinguish between these scenarios, we construct a 
weighted average, namely: $m(\hat{\rho})r_0= 5.9(6)$. 

 We note that the result of previous quenched calculations  was
summarised~\cite{Michael:2003ai} at around 1.9 GeV which corresponds to
$m(\hat{\rho})r_0=4.8$.
 Thus we conclude that the $N_f=2$ studies, using a fully unitary
theory,  tend to give higher hybrid meson masses than quenched. Our
averaged mass corresponds to 2.2(2) GeV (from $r_0=0.53$ fm). As
discussed later, there  are additional systematic errors.


\begin{figure}[t]
  \centering
    \includegraphics[width=1.2\linewidth]      {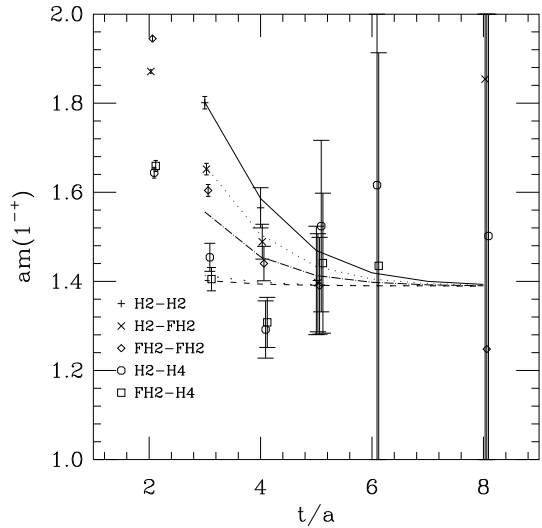}
        \caption{Effective mass (from $t:(t-a)$) of the $J^{PC}=1^{-+}$
hybrid meson in  lattice units from a $2\times 3$ matrix of
correlators for U355. The fit described in the text is illustrated. }
      \label{fig:hyb}
      \end{figure}

 We also have measured correlators between two-body operators and  the
hybrid meson - these we will discuss in the context of the decay
mechanism. They can also be used to extract energy values, as we discuss
later. First we analyse the reliability of the evaluation  of decay
transition strengths for S-wave decays at threshold by considering  a
case where the result is known experimentally: $b_1 \to \omega \pi$.

\begin{table}

\begin{tabular}{lllllll}

Code &  no. & $\kappa$ & $m(\pi)r_0$ & $L/r_0$ & 
 $am(\pi)$ & $am(\rho)$ \\
\hline \\

U355 & 200 & 0.1355 & 1.47 & 3.2 & 0.292(2) & 0.491(7) \\
C410 & 237 & 0.1410 & 1.29 & 5.3 & 0.427(1) & 0.734(4)\\
U395 12  & 20 & 0.1395   & 1.92 & 3.5  &0.558(8) &0.786(8)\\
U395 16  & 20 &0.1395    & 1.94 & 4.65 &0.564(3) &0.785(8)\\
\end{tabular}

\caption{Lattice gauge configurations U355 from
UKQCD~\cite{Allton:2001sk}, C410 from CP-PACS~\cite{AliKhan:2001tx} and
U395 from UKQCD~\cite{Allton:1998gi}  used.  These have $N_f=2$
flavours of sea quark  and we use valence quarks  of the same mass as
the sea quarks. 
 } 
 \label{tab1}
\end{table}

\begin{table}

\begin{tabular}{llll}

Code &  $am(b_1)$& $am(a_1)$ & $am(\hat{\rho})$\\
\hline \\

U355  & 0.78(2) &    0.72(2) & 1.39(6) \\
C410 & 1.17(3) &   1.15(2) & 1.78(5) \\  

\end{tabular}

\caption{ Axial meson and hybrid meson masses determined from lattice gauge
configurations U355 and C410 using stochastic methods.
 } 
 \label{tab2}
\end{table}

\subsection{Transition $b_1 \to \omega \pi$}

 The study of decays  from Euclidean lattice gauge theory has a  long
history and not many results. The only case that has been studied  in
detail on a lattice~\cite{McNeile:2002fh} where experimental results
exist is $\rho \to \pi \pi$ which has a P-wave decay, necessitating
non-zero momentum on a lattice.  For an S-wave decay, a zero momentum
transition is accessible and this we  explore here.
 
 On a lattice, one can create two body states of a given total momentum.
In  a large spatial volume, these two bodies will interact very little
and  so the two body state will be approximately the product of the
single body  states. The interaction can be studied by measuring the
shift in the  two-body energy while varying the lattice spatial size, as
 established  by
L\"uscher~\cite{Luscher:1986a,Luscher:1986b,Luscher:1990ux,Luscher:1991cf}.
This approach, however, needs very  accurate energy determinations and
is thus not feasible in many cases  at present. A simpler, but less
rigorous,  alternative is to measure the
 transition strength from initial state to two-body state directly  on a
lattice. This is feasible~\cite{McNeile:2002az}  when the initial and
final state have approximately the same energies (i.e. on-shell
transition). The decay transition amplitude measured on  a 
lattice can then be related to the large volume value via Fermi's golden 
rule. 
 For an S-wave decay at threshold, the phase space is actually zero, so 
relating the lattice with a discrete spectrum of two-body states to the 
large volume continuum of two-body states needs to be validated.

 Here we study $b_1 \to \omega \pi$ which has predominantly an S-wave 
decay~\cite{Eidelman:2004wy} with partial width $\Gamma= 0.142(9)$ GeV. 
We shall compare effective coupling constants for the S-wave transition,
defining $\overline{g}^2  = \Gamma/k$ where $k$ is the decay momentum,
so $\overline{g}^2=0.38(2)$. In order to study the $b_1 \to \omega \pi$
D-wave  transition which would shed light on the  decay
mechanism~\cite{Barnes:2003bn}, we would have to create non-zero
momentum mesons which we do not study in this work.

 On our lattices, the $b_1$ mass is approximately the same as the sum of
 $\pi$ and $\rho$ masses, so we are close to  an on-shell transition. In
principle, the $\omega$ meson, which is  flavour singlet, receives 
disconnected contributions, but these we expect~\cite{McNeile:2001cr} to
be negligible based on previous  lattice studies. We measure the
three-point correlation $(b_1 | \omega \pi)$ where the $\pi$ and
$\omega$ are  both summed over the whole volume at one time-slice and
the $b_1$ meson  is also summed over all volume at another time-slice.
We measure  for each of  the three spin  orientations of $b_1$ and
$\omega$ (which are aligned). This is achieved using  our stochastic
methods with the $\omega$ as the source and with an extended propagator
for the  zero-momentum pion, for which a fuzzed operator is chosen.

 We then evaluate the ratio
 $$
  R(t)= { (b_1 | \omega \pi) \over
   \sqrt{(b_1 | b_1) (\omega | \omega)(\pi | \pi)} }
    $$
 where each correlation is taken at the same $t$-separation on the
lattice. This ratio $R(t)$ of the three point correlation to a
combination  of two-point correlations normalises the meson creation
operators. It is plotted  in fig.~2 for the case where each meson
operator is fuzzed (which enhances the ground state contributions). The 
slope of $R$ versus $t$ is then the lattice transition amplitude $xa$.
We see that a linear behaviour is present over a  significant interval
in $t$, so confirming the  interpretation of the slope as the lattice
transition amplitude. In our further analysis, we evaluate this slope
around $t/r_0=1$. 
 The decay width $\Gamma$ is then given, via Fermi's Golden Rule,  by
$\Gamma/k=\overline{g}^2$, where 
$$
 \overline{g}^2 = {1 \over  \pi} (xa)^2 (L/a)^3 
   { E(\omega)a E(\pi) \over E(\omega)+E(\pi) }
 $$
 where $k$ is the centre of mass momentum of the decay products.

\begin{figure}[t]
  \centering
    \includegraphics[width=1.2\linewidth]      {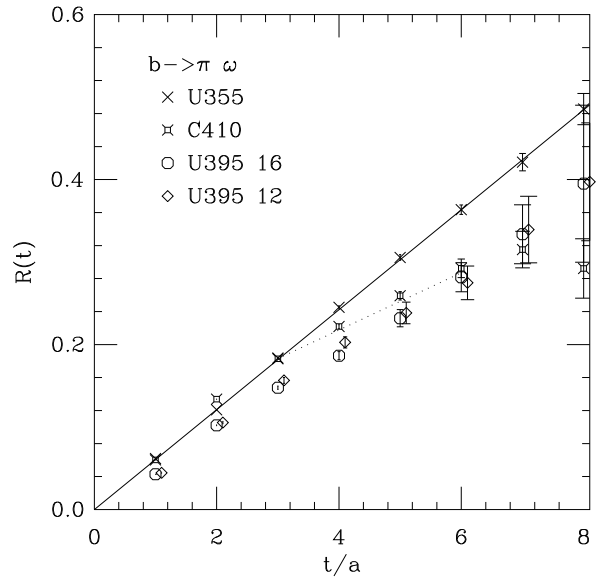}
        \caption{ Normalised ratio $R(t)$ of transition $b_1 \to \pi
\omega$  at time $t$ in lattice units (a factor of $(12/16)^{3/2}$ has
been  included for the U395 12 data set). The continuous (dotted)
straight lines represent fits to the  expected behaviour for U355 (C410)
and their slopes ($xa$) are related to the effective coupling constant
$\overline{g}$ as described in the text.  }
      \label{fig:xtb}
      \end{figure}

\begin{figure}[t]
  \centering
    \includegraphics[width=1.2\linewidth]      {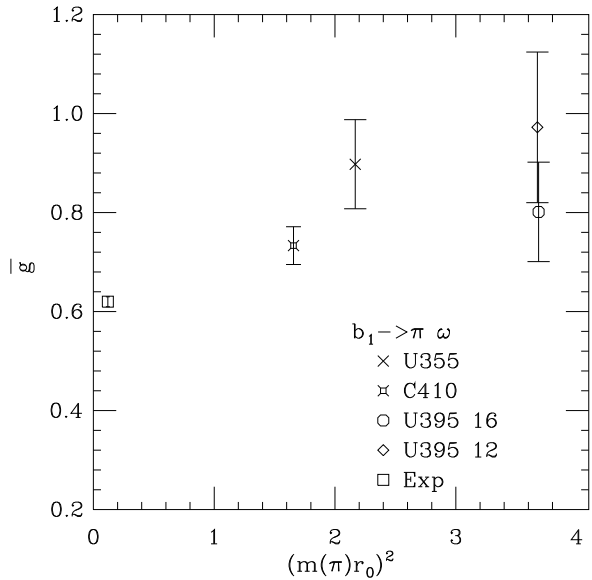}
        \caption{ Effective coupling of transition $b_1 \to \pi
\omega$  evaluated from the slope in fig.~2 at time $t/r_0=1$ for different 
lattices with different pion masses (in units of $r_0 \approx 0.5$ fm). 
The strange quark mass corresponds to $(m(\pi)r_0)^2 \approx 3.4$.
 }   
      \label{fig:slope}
      \end{figure}

 The coupling strength $\overline{g}^2$ should be independent of  lattice
spatial size ($L$). Moreover its dependence on the  quark mass can also
be explored. In order to investigate, we evaluate  the effective
coupling from a range of different lattice configurations (all having 
$N_f=2$ flavours of sea-quark). 
   We plot in fig.~3  the  effective coupling $\overline{g}$ evaluated 
from the slope ($xa$) in fig.~2 at $t/r_0=1$.

For a  zero-momentum  transition, which is what we study here, there
should be  no dependence on lattice volume of the hadronic transition
strength.
 In order to be able to vary the lattice spatial volume while keeping
everything else fixed, we make use of $12^3 \times 24$ and $16^3 \times
24$ configurations  labelled U395 in Table~\ref{tab1}. As shown on fig.~2, where
the appropriate factor of $(12/16)^{3/2}$  has been included,   $R(t)$ 
agrees within statistical errors for these two cases. This confirms that
the extraction of the  hadronic transition amplitude is insensitive to
the lattice spatial volume when  it is changed by a factor of 2.4.

To study the dependence on the quark mass, we compare our higher
statistics studies U355 ($m(\pi)r_0= 1.47$) and C410 ($m(\pi)r_0=
1.29$). Qualitatively, we see from fig.~2, they have similar  transition
strengths. There is some evidence of a decrease of  the coupling
strength as the quark mass is decreased. This is consistent with
approaching the experimental value   as $m(\pi) \to 0$, as shown in
fig.~3.


Thus we conclude that our method for extracting an estimate of the 
decay transition strength is indeed reliable for an S-wave decay 
at threshold. We now explore hybrid meson decays.

\subsection{Hybrid meson decay}

With our lattice parameters, there are several two-body thresholds with 
the quantum numbers of the hybrid meson   in the energy range
of interest, namely $\pi f_1$, $\pi b_1$,  $\pi \eta$ and $\pi \rho$,
where the latter two cases involve non-zero momentum  since they are
P-wave decays. 
 We first investigate the coupling between the hybrid meson $\hat{\rho}$
 and the two-body channel $\pi b_1$. This latter channel has an S-wave
coupling  to the hybrid, so the energy threshold on our lattice is given
by  the values of $aE(\pi,b_1)$ in Table~\ref{tab3}. These values
are similar to our estimate of the hybrid meson mass,    so that the
normalised off-diagonal transition gives useful information.  We use
similar methods for $\hat{\rho} \to b_1 \pi$ as used above  for $b_1 \to
\pi \omega$.

The simplest way to investigate the hadronic matrix element responsible for 
decay $\hat{\rho} \to b_1 \pi$ is to evaluate the ratio 
 $$
 H(t)= { (\hat{\rho} | b_1 \pi) \over
 \sqrt{(\hat{\rho} | \hat{\rho}) (b_1 | b_1)(\pi | \pi)} }
 $$
 where each correlation is taken at the same $t$-separation on the
lattice. This ratio normalises the meson creation operators. One could 
also normalise directly the two-particle state ($\pi b_1$) but,  as
discussed~\cite{McNeile:2002fh} for $\rho \to \pi \pi$,  we expect the
dominant contribution to be the product (especially for large $L$).
Indeed we  do check that the correlation between these  two-point
correlators  $(\pi | \pi)$ and $(b_1 | b_1)$ is consistent with zero
out to $t=8a$.

If there were only a single state coupling to the $\hat{\rho}$
and $\pi b_1$  operators, the ratio $H$ would be 1.0. The observed value
is small (see fig.~4) which shows that this  is not the case.

\begin{figure}[t]
  \centering
    \includegraphics[width=1.2\linewidth]      {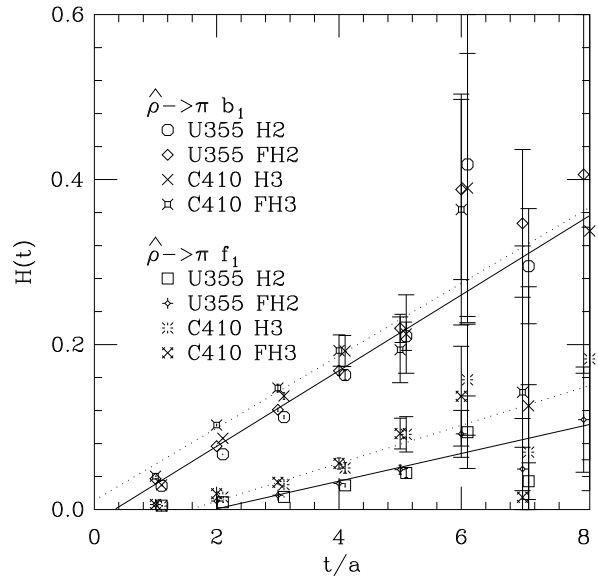}
        \caption{Normalised  ratio of $\hat{\rho} \to b_1 \pi$ and 
$\hat{\rho} \to f_1 \pi$. The operator used for $\hat{\rho}$ is  either
U-bends of size 2 or 3 (H2 or H3) or the same combined with fuzzing (FH2
or FH3).  The $\pi$ and axial meson are always fuzzed. The straight
lines represent the trend of the data: the slope ($xa$) is the quantity
required. 
 }
      \label{fig:xt}
      \end{figure}

\begin{table}

\begin{tabular}{llllll}

Code & $aE(\pi,b_1)$ & $\hat{\rho} \to b_1 \pi$ && $\hat{\rho} \to f_1 \pi$ & \\ 
\hline 
 & & $xa$ & $\Gamma/k$ & $xa$ & $\Gamma/k$ \\
\hline 
U355 & 1.07(2) & 0.046(5) & 0.58(12) & 0.017(7) & 0.08(6) \\
C410 & 1.60(4) & 0.045(7) & 0.82(26) & 0.023(7) & 0.22(13)\\
Average &    &            & 0.66(20) &          & 0.15(10) \\
\end{tabular}


\caption{Hybrid meson decay amplitudes and rates.
 } 

 \label{tab3}

\end{table}

In the case where there are two states (the hybrid meson and the
two-body threshold), the ratio $H(t)$ then behaves~\cite{McNeile:2002az}
as $xt$ where $x$ is the lattice transition amplitude,  provided that
the transition is approximately on-shell and  that $xt << 1$. We do
indeed observe an  approximately linear behaviour and, moreover, the
value  is consistent between different choices of external operator
(fuzzed or not)  for both the $b_1$ meson and the $\hat{\rho}$.   Thus
we can assume that  the ground state mesons dominate and read off the 
lattice transition amplitudes which are given in Table~\ref{tab3}.
Statistically these slopes are quite well determined, although as we
discuss later, the systematic errors  are dominant.  
 The decay width $\Gamma$ is given, via Fermi's Golden Rule,  by
 $$
 \Gamma/k = {1 \over  \pi} (xa)^2 (L/a)^3 
   { E(b_1)a E(\pi) \over E(b_1)+E(\pi) }
 $$
 where $k$ is the centre of mass momentum of the decay products.

 Then using our observed mass values for the $\pi$ and $b_1$,  the
values we obtain for the partial width $\Gamma/k$  of the decay
$\hat{\rho} \to b_1 \pi$ are given in Table~\ref{tab3}. We emphasise that, on
the lattice,  we are working with unphysical light quark masses  which
makes the transition  nearly on-shell.
 The underlying assumption, however, is that the coupling constant $g$
(where  $\Gamma/k$ is an effective proxy for $g^2$) is  insensitive to
changes in the quark masses. This was confirmed by our
study~\cite{McNeile:2002fh} of $\rho$ decay and our study, above, of
$b_1$  decay shows only a relatively small dependence (see fig.~3).

As shown in Table~\ref{tab3}, the resulting values of the coupling (quoted as
$\Gamma/k$) vary between the two lattice evaluations we have used. As
discussed  above for $b_1$ decay, each set of lattice configurations has
 favourable and less favourable features (large volume, smaller lattice
spacing, lighter quarks, etc.). The best way forward is to regard these 
two studies  as indicative of the systematic errors arising from these 
limitations. As a compromise we quote averaged effective couplings in
Table~\ref{tab3} which take into account some of these systematic
errors.

As well as $\hat{\rho} \to \pi b_1$, we can also explore another S-wave
decay: $\hat{\rho} \to \pi f_1$.  In this case, a disconnected diagram
also  contributes to the decay, but we expect the contribution from 
such disconnected diagrams to be small for the axial-vector
meson~\cite{McNeile:2001cr}. We also assume that the $f_1$ meson,
because of  the small disconnected contribution, is dominantly
$\bar{u}u+ \bar{d}d$.  Then the relevant ratio is shown also in  fig.~4.
This again shows a linear increase with $t$, leading to  estimates of 
the hadronic transition $xa$  and partial width  $\Gamma/k$ for this
decay shown in Table~\ref{tab3}. These substantially smaller values are only 
partly attributable (a factor 0.5 in rate) to quark diagram counting 
and are dominantly a dynamical effect.

If one was convinced that the disconnected contributions could be 
neglected for $\hat{\omega}$ decays to  $\pi a_1$ then our 
estimate for $x$ obtained above can be used. With the same assumptions
we  would obtain a partial width $\Gamma/k$  of 0.45(30).


The excited two body state where  $\pi$ and $b_1$ have momenta $\pm 2
\pi/L$  has an energy estimated on our U355 lattice at $Ea=1.36$. This
is close to the energy  we find for our hybrid state. It would be
desirable to evaluate the transition from $\hat{\rho}$ to  this excited
two-body state to check for a consistent estimate of the decay width. On
a lattice, however, this  excited two-body state will be very hard to
isolate because of the dominant contribution  of the threshold state
which has the same overall quantum numbers.

In this work we do not explore the P-wave decay channels such as 
$\pi \rho$; $\pi \eta$ etc., since we have not introduced non-zero 
momentum, although as in the case of $\rho$ decay, this is in principle 
possible.

With $N_f=2$ flavours of sea-quark, we expect the $\hat{\rho}$ meson to 
be mixed with the two body channels such as $\pi b_1$. Indeed by 
careful measurement of the energy of the two-body state, it is possible
to
deduce~\cite{Luscher:1986a,Luscher:1986b,Luscher:1990ux,Luscher:1991cf} 
the scattering phase shift and, hence, properties of the $\hat{\rho}$
resonance, if there is only one two-body channel open. 
 In this case, however, several channels are open which invalidates the
assumptions. Furthermore, we do not have sufficient precision in our
energy determination  to pursue this although preliminary attempts have
been made~\cite{Cook:2005az}.

 We can, however, attempt a joint fit to the matrix of operators
available:  three $\hat{\rho}$ operators (as discussed above) and also
the  $\pi b_1$ operator where the $\pi$ is fuzzed but the $b_1$ may be
either  local or fuzzed. We thus have a $4 \times 5$ matrix of
correlations, assuming that   $(\pi b_1 | \pi b_1)$ can be replaced by
the product $(\pi | \pi)(b_1 | b_1)$  which we do measure. For U355,
from such a 3-state fit for the t-range 4-8 we obtain lowest energies of
 $0.93(15)/a$ and $1.41(8)/a$. This is indeed consistent with the
picture  we have assumed so far.


\section{Discussion}

 We study the spin-exotic channel with $J^{PC}=1^{-+}$ and we do obtain a
signal for a state additional to the two-body threshold. On our lattices
this state is  relatively heavy - at 2.2(2) GeV.
 We  find consistent mass values from the  hybrid channel alone and when
the $\pi b_1$ channel is included. 
 Since mixing between the discrete two-body channels and the hybrid
meson  is enabled on our lattice, it is possible that this mixing  moves
the hybrid mass up - but this shift would be expected to be  of order
$xa$ which we find to be only 0.05 and hence within our quoted
statistical error. 
   As we only have a signal out to $t/a \approx 6$, we will not be
able to resolve  a rich hybrid spectrum - since we are only able to make
2-state fits to the  hybrid sector. So it is possible that there are
several hybrid mesons in this mass  region, one of which is lighter than
our mass estimate. Indeed our variational estimate explicitly is an
upper mass  estimate. We do have some control over the contribution of
two-body states to  the hybrid sector and those we explicitly measure
contribute  only a negligible amount. 

 For our quark masses (approximately 2/3 of strange for U355 and 1/2 for
C410), previous lattice  results gave a lighter hybrid mass (around 2
GeV), but they were predominantly quenched.  We do not attempt a mass
extrapolation,  although phenomenological
estimates~\cite{Lacock:1996vy,Lacock:1996ny} would be that the
light-quark hybrid state is  some 200 MeV lighter than that we study
here. For further discussion of the subtleties of extrapolating 
in quark mass see ref.~\cite{Thomas:2001gu,Hedditch:2005zf}.
 We are also unable to extrapolate to smaller lattice spacing or to 
larger spatial volumes. The neglect of strange sea-quarks is also  hard
to quantify but could be quite small for the states we consider.
Even though these restrictions imply that our mass determination has 
systematic errors, we are considering a consistent lattice field theory 
and we expect hadronic transition strengths to be a good indication of the 
physical world.

 The signal for the hadronic transition which causes decay is very 
clear and is consistent with the simple interpretation in which the
slope  gives the decay amplitude. The most suitable way to quote our
result is  as effective couplings given by partial widths
$\Gamma(\hat{\rho} \to \pi b_1)/k = 0.66(20)$ and  $\Gamma(\hat{\rho}
\to \pi f_1)/k = 0.15(10)$. These error estimates do not include any
error from extrapolation  to physical quark masses.  If the $\hat{\rho}$
meson is  at 2.0 GeV then the physical decay to $\pi b_1$ has momentum
$k=0.611$ MeV  and for a heavier hybrid meson the momentum would be even
larger. This implies that we  expect the hybrid meson to have a  partial
width  to $\pi b_1$ of  400(120) MeV which implies  a total width
greater than this.

There are experimental indications for a $\hat{\rho}$ resonance around 2
GeV with a  total decay width reported as  333 MeV (from
ref.~\cite{Kuhn:2004en} studying $\pi f_1$ final states)  and as 230 MeV
(from ref.~\cite{Lu:2004yn} studying $\pi b_1$ final states). These
total width values should be the same if there is one underlying state
and can be compared with the total width we estimate.  The agreement is
close enough  to warrant further experimental investigation. It would be
of considerable interest to know if the  experimental branching
fractions (as yet unknown)  tie in with our expectation (namely
dominance of $\pi b_1$ over $\pi f_1$). Indeed phenomenological  models
do indicate~\cite{Thomas:2001gu} that the $\pi b_1$ mode should dominate
the width  and that widths of $O(100)$ MeV are possible. Flux tube
models~\cite{Isgur:1985vy,Close:1994hc} give $\Gamma/k$ in the range .06
to .28 for $\pi b_1$ and .04 to .10 for  $\pi f_1$.

 The equivalent decay to $\hat{\rho} \to \pi b_1$ for a  heavy-quark
hybrid will be $\hat{\Upsilon} \to B(0^-) B^{**}(1^+)$  which is not
expected to be allowed energetically. So the previous
estimate~\cite{McNeile:2002az} for the  decay width of the 
$\hat{\Upsilon}$ is not modified by this work.

\section{Conclusions}

We have evaluated the S-wave transition $b_1 \to \pi \omega$ at
threshold  from the lattice and obtained agreement with the experimental
value. We find  some evidence that the coupling constant varies with the
quark mass, being smaller  for lighter quarks.

We have studied the spectrum and decay to $\pi b_1$ and $\pi f_1$  of
the spin exotic isovector hybrid meson $\hat{\rho}$. This state has 
potential couplings to many two-body states in the same energy region 
which inevitably means that approximate methods will be needed. We find
 statistically well determined results in our study which are consistent
with a  hadronic transition from $\hat{\rho}$ to $\pi b_1$ and $\pi f_1$
whose  strength we evaluate. From this lattice determination, assuming
that the  effective coupling constant is independent of quark mass, one
can estimate the  physical partial widths, obtaining $\Gamma(\hat{\rho}
\to \pi b_1)/k=0.66(20)$ and $\Gamma(\hat{\rho} \to \pi f_1)/k=0.15(10)$
 where $k$ is the decay momentum. Note that, if the result we found for
$b_1 \to \pi \omega$ is generic and  the effective coupling decreases
with quark mass, then the  physical decay width of  the hybrid meson
would be smaller than our estimates. We note that the width of the 
hybrid meson is large primarily because of larger phase space rather
than larger  coupling, compared to  the decay of a typical meson, such
as $b_1 \to \pi \omega$.

 Our determination of the mass  of this hybrid meson gave higher values
than obtained previously  for our lattice with smaller lattice spacing
(U355) but averaging over our  two sets of lattices we obtain 2.2(2) GeV
for light quarks of similar mass to strange, which is similar to the
value  2.0(2) GeV previously determined. Because of the possibility of a
rich spectrum  (both of hybrid mesons and of two-body states) we cannot
exclude systematic errors  in our mass determination and we can only be
certain that it is an upper limit.  We do, however, see some evidence
that  the hybrid meson may lie higher in energy when the two-body decay
channels are  coupled (as they are with dynamical sea quarks). 

The study of the properties of an unstable state (namely the spin-exotic
hybrid meson)  demands careful treatment on a lattice. We have shown
that this is feasible and   future studies with lattices closer 
to the continuum and with lighter quarks and  higher statistics will allow 
further refinement of our estimates.

 Overall our results are rather disappointing from a viewpoint of
experimental  searches for spin-exotic hybrid mesons. We have evaluated
two decay channels  which combine to give a total width of over 400 MeV.
This will make the  detailed experimental study of the hybrid meson
relatively difficult. There are detailed predictions (such as that the 
$\pi b_1$ mode will be dominant) that can be checked, however.

\section{Acknowledgements}
 
One of the authors (CM) wishes to thank PPARC for the award of a Senior
Fellowship.
 This work has been supported in part by the EU Integrated
 Infrastructure Initiative Hadron Physics (I3HP) under contract
  RII3-CT-2004-506078.
 We acknowledge the ULGrid project of the University of Liverpool for making
 available  computer resources.
 We acknowledge the CP-PACS collaboration~\cite{AliKhan:2001tx} for
making available  their gauge  configurations. 


\begin{thebibliography}{10}

\bibitem{McNeile:2002az}
UKQCD, C.~McNeile, C.~Michael, and P.~Pennanen,
\newblock Phys. Rev. {\bf D65}, 094505 (2002), hep-lat/0201006,
\newblock 

\bibitem{Lacock:1996vy}
UKQCD, P.~Lacock, C.~Michael, P.~Boyle, and P.~Rowland,
\newblock Phys. Rev. {\bf D54}, 6997 (1996), hep-lat/9605025,
\newblock 

\bibitem{Lacock:1996ny}
UKQCD, P.~Lacock, C.~Michael, P.~Boyle, and P.~Rowland,
\newblock Phys. Lett. {\bf B401}, 308 (1997), hep-lat/9611011,
\newblock 

\bibitem{Bernard:1997ib}
MILC, C.~W. Bernard {\em et~al.},
\newblock Phys. Rev. {\bf D56}, 7039 (1997), hep-lat/9707008,
\newblock 

\bibitem{Hedditch:2005zf}
J.~N. Hedditch {\em et~al.},
\newblock Phys. Rev. {\bf D72}, 114507 (2005), hep-lat/0509106,
\newblock 

\bibitem{Lacock:1998be}
TXL, P.~Lacock and K.~Schilling,
\newblock Nucl. Phys. Proc. Suppl. {\bf 73}, 261 (1999), hep-lat/9809022,
\newblock 

\bibitem{Bernard:2003jd}
C.~Bernard {\em et~al.},
\newblock Phys. Rev. {\bf D68}, 074505 (2003), hep-lat/0301024,
\newblock 

\bibitem{Alde:1988bv}
IHEP-Brussels-Los Alamos-Annecy(LAPP), D.~Alde {\em et~al.},
\newblock Phys. Lett. {\bf B205}, 397 (1988),
\newblock 

\bibitem{Thompson:1997bs}
E852, D.~R. Thompson {\em et~al.},
\newblock Phys. Rev. Lett. {\bf 79}, 1630 (1997), hep-ex/9705011,
\newblock 

\bibitem{Abele:1998gn}
Crystal Barrel, A.~Abele {\em et~al.},
\newblock Phys. Lett. {\bf B423}, 175 (1998),
\newblock 

\bibitem{Adams:1998ff}
E852, G.~S. Adams {\em et~al.},
\newblock Phys. Rev. Lett. {\bf 81}, 5760 (1998),
\newblock 

\bibitem{Chung:1999we}
E852, S.~U. Chung {\em et~al.},
\newblock Phys. Rev. {\bf D60}, 092001 (1999), hep-ex/9902003,
\newblock 

\bibitem{Ivanov:2001rv}
E852, E.~I. Ivanov {\em et~al.},
\newblock Phys. Rev. Lett. {\bf 86}, 3977 (2001), hep-ex/0101058,
\newblock 

\bibitem{Dzierba:2005jg}
A.~R. Dzierba {\em et~al.},
\newblock (2005), hep-ex/0510068,
\newblock 

\bibitem{Kuhn:2004en}
E852, J.~Kuhn {\em et~al.},
\newblock Phys. Lett. {\bf B595}, 109 (2004), hep-ex/0401004,
\newblock 

\bibitem{Lu:2004yn}
E852, M.~Lu {\em et~al.},
\newblock Phys. Rev. Lett. {\bf 94}, 032002 (2005), hep-ex/0405044,
\newblock 

\bibitem{Michael:2005kw}
C.~Michael,
\newblock PoS {\bf LAT2005}, 008 (2005), hep-lat/0509023,
\newblock 

\bibitem{McNeile:2002fh}
UKQCD, C.~McNeile and C.~Michael,
\newblock Phys. Lett. {\bf B556}, 177 (2003), hep-lat/0212020,
\newblock 

\bibitem{Foster:1998vw}
UKQCD, M.~Foster and C.~Michael,
\newblock Phys. Rev. {\bf D59}, 074503 (1999), hep-lat/9810021,
\newblock 

\bibitem{Foley:2005ac}
J.~Foley {\em et~al.},
\newblock Comput. Phys. Commun. {\bf 172}, 145 (2005), hep-lat/0505023,
\newblock 

\bibitem{Lacock:1994qx}
UKQCD, P.~Lacock, A.~McKerrell, C.~Michael, I.~M. Stopher, and P.~W.
  Stephenson,
\newblock Phys. Rev. {\bf D51}, 6403 (1995), hep-lat/9412079,
\newblock 

\bibitem{Michael:2003ai}
C.~Michael,
\newblock Int. Rev. Nucl. Phys. {\bf 9}, 103 (2003), hep-lat/0302001,
\newblock 

\bibitem{Allton:2001sk}
UKQCD, C.~R. Allton {\em et~al.},
\newblock Phys. Rev. {\bf D65}, 054502 (2002), hep-lat/0107021,
\newblock 

\bibitem{AliKhan:2001tx}
CP-PACS, A.~Ali~Khan {\em et~al.},
\newblock Phys. Rev. {\bf D65}, 054505 (2002), hep-lat/0105015,
\newblock 

\bibitem{Allton:1998gi}
UKQCD, C.~R. Allton {\em et~al.},
\newblock Phys. Rev. {\bf D60}, 034507 (1999), hep-lat/9808016,
\newblock 

\bibitem{Luscher:1986a}
M.~Luscher,
\newblock Commun. Math. Phys. {\bf 104}, 177 (1986).

\bibitem{Luscher:1986b}
M.~Luscher,
\newblock Commun. Math. Phys. {\bf 105}, 153 (1986).

\bibitem{Luscher:1990ux}
M.~Luscher,
\newblock Nucl. Phys. {\bf B354}, 531 (1991),
\newblock 

\bibitem{Luscher:1991cf}
M.~Luscher,
\newblock Nucl. Phys. {\bf B364}, 237 (1991),
\newblock 

\bibitem{Eidelman:2004wy}
Particle Data Group, S.~Eidelman {\em et~al.},
\newblock Phys. Lett. {\bf B592}, 1 (2004),
\newblock 

\bibitem{Barnes:2003bn}
T.~Barnes,
\newblock AIP Conf. Proc. {\bf 717}, 625 (2004), hep-ph/0311102,
\newblock 

\bibitem{McNeile:2001cr}
UKQCD, C.~McNeile, C.~Michael, and K.~J. Sharkey,
\newblock Phys. Rev. {\bf D65}, 014508 (2002), hep-lat/0107003,
\newblock 

\bibitem{Cook:2005az}
M.~S. Cook and H.~R. Fiebig,
\newblock PoS {\bf LAT2005}, 062 (2005), hep-lat/0509028,
\newblock 

\bibitem{Thomas:2001gu}
A.~W. Thomas and A.~P. Szczepaniak,
\newblock Phys. Lett. {\bf B526}, 72 (2002), hep-ph/0106080,
\newblock 

\bibitem{Isgur:1985vy}
N.~Isgur, R.~Kokoski, and J.~Paton,
\newblock Phys. Rev. Lett. {\bf 54}, 869 (1985),
\newblock 

\bibitem{Close:1994hc}
F.~E. Close and P.~R. Page,
\newblock Nucl. Phys. {\bf B443}, 233 (1995), hep-ph/9411301,
\newblock 


\end{thebibliography}

\end{document}